\documentstyle[11pt]{article}
\begin{document}
\renewcommand{\thefootnote}{\fnsymbol{footnote}}
\newcommand{\fracts}[2]{\textstyle \frac{#1}{#2} }
\newcommand{\fracds}[2]{\displaystyle \frac{#1}{#2} }
\newcommand{\eq}[1]{eq.(\ref{#1})}
%
\renewcommand{\theequation}{\arabic{section}.\arabic{equation}}
\newcommand{\sectione}[1]{ \section{#1} \setcounter{equation}{0} }
%
\newcommand{\startapp}{ \appendix
\renewcommand{\theequation}{\Alph{section}.\arabic{equation}} }
\newcommand{\startappA}{ \setcounter{equation}{0} \appendix
\renewcommand{\theequation}{A.\arabic{equation}} }
\newcommand{\bea}{\begin{eqnarray}}
\newcommand{\eea}{\end{eqnarray}}
\def\be{\begin{equation}}
\def\ee{\end{equation}}
\def\qexp{Exp_{q^2}}
\def\bqexp{Exp_{q^{-2}}}
\newcommand{\zerosection}{\setcounter{section}{-1}}

\def\sections#1#2{\bigskip\noindent{\bf #1\quad #2}\bigskip}
\def\proof{\medskip\noindent{\sl Proof\ } :\  }
\def\theorem#1{\medskip\noindent{\bf Theorem\ } #1 :\ }
\def\condition#1
{\medskip\noindent{\bf Condition\ } #1 :\ }
\def\corollary#1{\medskip\noindent{\bf Corollary\ } #1 :\ }
\def\proposition#1{\medskip\noindent{\bf Proposition\ } #1 :\ }
\def\definition#1{\medskip\noindent{\bf Definition\ } #1 :\ }
\def\question{\noindent{bf Question\ } :\ }
\def\qed{\rightline{\sl q.e.d}\medskip}


\def\and{\quad{\rm and}\quad}
\def\OR{\quad{\rm or}\quad}
\def\for{\quad{\rm for}\quad}
\def\where{\quad{\rm where}\quad}
\def\with{\quad{\rm with}\quad}
\def\Det{{\rm det}}
\def\Dirac#1{ #1 \hskip-6pt /\,}
\def\half{{1\over2}}
\def\p{\partial}
\def\Tr#1{{\rm T}\!{\rm r}\{#1\}}
\def\tr{{\rm T}\!{\rm r}}
\def\wed{\mathop{\wedge}}
\def\intdx#1{\int\! d^{#1}x\ }
\def\brs{\delta^{\ }_B{}}
\def\one{{\bf1}}

\def\hs{\hat{s}}
\def\R{{\widehat{\bf R}}}
\def\M{M{}}
\def\opst{{\bf p}}
\def\cp{{\cal P}}
\def\cpa#1{{\cal P}_A\big(#1\big)}

\def\al{{\cal A}}
\def\qal{{\cal M}}
\def\co{\Delta}
\def\col{\Delta_L}
\def\cor{\Delta_R}
\def\I{{\bf I}}
\def\tchi{\widetilde{\chi}}
\def\e{{\varepsilon}}
\def\A{{\bf A}}
\def\tA{\widetilde{\bf A}}
\def\btheta{\bar{\theta}}
\def\C{{\rm C}}
\def\X{{\bf X}}
\def\d{{\bf d}}
\def\norm{{\cal N}}
\def\antipode#1{\kappa\big(#1\big)}
\def\ap{S}
\def\bra{[\![}
\def\ket{]\!]}
\def\anti{\kappa}
\def\cs{{\cal S}}

\begin{titlepage}
\vspace*{-4ex}


\null \hfill Preprint TU-442  \\
\null \hfill August, 1993 \\[4ex]

\begin{center}

\bigskip
\bigskip

{\Large \bf
The $q$-deformed Schr\"odinger Equation of  the
}\\[3ex]
{\Large\bf
Harmonic Oscillator on the Quantum Euclidian Space
}\\ [18ex]

  Ursula Carow-Watamura \  and \  Satoshi Watamura

    Department of Physics \\
 Faculty of Science \\
 Tohoku University \\
 Aoba-ku, Sendai 980, JAPAN \\ [2ex]
 \end{center}

\bigskip
\begin{abstract}
\medskip
We consider the $q$-deformed Schr\"odinger equation of the harmonic
oscillator on the $N$-dimensional quantum Euclidian space.
The creation and annihilation operator are found, which systematically
produce all energy levels and eigenfunctions of the Schr\"odinger
equation. In order to get the $q$-series representation of the
eigenfunction,
 we also give an alternative way to solve the Schr\"odinger equation
 which is based on the $q$-analysis.  We represent the Schr\"odinger
 equation  by the $q$-difference
equation and solve it by using $q$-polynomials and $q$-exponential
functions.
 \end{abstract}
 \end{titlepage}

\sectione{Introduction}

\bigskip

To investigate the possibility of defining the quantum mechanics on a
different type of geometry is a very interesting problem since we can expect
that it will show new aspects of the quantum theory, see for example
\cite{[Pod]}.
When we consider the noncommutative geometry as the base geometry, one
feature is that the theory is formulated in an algebraic language
since the noncommutative space is defined by a function algebra.
Therefore, it is not easy to use the analogy with the commutative geometry
in a simple way through such
an algebraic language.

Taking the point of view that the quantum group and
the quantum space are the $q$-deformation of the usual group and space, we
can find the noncommutative analogue of the known objects such as the
quantum
Lorentz group \cite{[Wor],[CSSW]},
the quantum Minkowski space \cite{[CSSW]}, the quantum Poincar\'e group
\cite{[Ruegg],[Wess],[ScWeWe],[Majid]} and many other properties.
Differential calculi on $q$-spaces have been also constructed in a rather
early stage of the investigations in quantum groups \cite{[WorPusz],[WZ]}.
They give a simple example of the noncommutative geometry and allow us to
draw the analogy to the non-deformed theory.

In ref.\cite{[CSW]} the authors have constructed the
differential calculus on the N-dimensional $q$-Euclidian space, i.e. the
differential calculus covariant under the action of the quantum
group $Fun_q(SO(N))$.  Although it became a little more complicated than the
one in refs.\cite{[WorPusz],[WZ]} which is based on $A$-type quantum groups,
it has the advantage that it contains the metric which makes it possible to
define the Laplacian.

 Using this differential calculus we have investigated the Schr\"odinger
 equation corresponding to the
$q$-deformed harmonic oscillator and have computed the ground state energy
as well as the first two excited energy levels.
However in that stage a systematic construction to all energy levels has
been missing.

Recently the investigations on this line were put forward by
several authors \cite{[Fio2],[Fio1],[CWHSW],[HW]}, especially by
the work of
Fiore \cite{[Fio1]}
who proposed raising and lowering operators which map
the wavefunctions of the $r$-th level to the one of the $(r+1)$th level.
However, these operators defined in ref.\cite{[Fio1]} have an explicit
dependence on the energy level, i.e., there is one such operator for each
level
and thus we have an infinite number of them. Due to this feature, these
operators cannot be considered as the $q$-analogue of the
creation-annihilation operator.

It is the aim of this paper to develop further this approach and to
construct the creation-annihilation operator of the N-dimensional
$q$-deformed harmonic oscillator which is level-independent.  This defines
all energy eigenvalues and gives
the expression of the eigenfunctions in terms of creation operators acting
on the ground state.  Since the creation-annihilation operator is given in
terms of differential operators, in principle all eigenfunctions can be
computed.
It is however not so easy to find the explicit form of the wave function as
a $q$-polynomial in the coordinates $x^i$ with this method.

Thus, we develop
an alternative method to construct the eigenfunctions of the
Schr\"odinger equation, which is another new result of this paper.
It corresponds to the analytic construction of the
wave function in the non-deformed case. Reducing the
Schr\"odinger equation to a $q$-difference equation we solve it directly by
using  $q$-polynomials. The above described two constructions give the same
eigenvalues and the resulting wave functions have a
one-to-one correspondence.

Since the investigations in this paper are based on the results obtained in
ref.\cite{[CSW]}, let us briefly recall them here.
It is known that the $\hat R$-matrix of the quantum group $Fun_q(SO(N))$ can
be written by using projection operators as
\begin{equation}
\hat R=q{\cal P}_S-q^{-1}{\cal P}_A +q^{1-N}{\cal P}_1 ,   \label{(I.1)}
\end{equation}
where the indices $S$, $A$ and $1$ denote symmetric, antisymmetric and
singlet, respectively.\cite{[FRT]}
The differential calculus on the $q$-Euclidian space derived in
ref.\cite{[CSW]} is making use of this projector decomposition. It is
defined by the algebra ${\bf C}\!<x^i, dx^i, \p^i>$ with relations
consistent with the quantum group action.  The relations are given by:

\begin{equation}
{\cal P}_A{}^{ij}_{kl}x^k\,x^l=0 ,  \label{(I.2)}
\end{equation}
\begin{equation}
{\cal P}_S(dx\wedge dx)=0\ ,\quad {\cal P}_1(dx\wedge dx)=0 \  ,
\label{(I.4)}
\end{equation}
\begin{equation}
x^idx^j=q\hat R^{ij}_{kl}dx^k x^l  ,  \label{(I.3)}
\end{equation}
\begin{equation}
\p^ix^j=C^{ij} +q\hat R^{-1 ij}_{\ \ kl}x^k\p^l  ,  \label{(I.5)}
\end{equation}
\begin{equation}
{\cal P}_{A\ kl}^{\ \ ij}\p^k\p^l=0   ,  \label{(I.6)}
\end{equation}
\begin{equation}
\p^idx^j=q^{-1}\hat R^{ij}_{kl}dx^k\p^l ,  \label{(I.7)}
\end{equation}
where $Q_N$ is a constant given by
\begin{equation}
Q_N={(1-q^N)\mu\over (1-q^2)}=C^{ij}C_{ij}     ,  \label{(I.9)}
\end{equation}
$C_{ij}$ is the metric of the quantum space,
and $\mu$ is
\begin{equation}
\mu=1+q^{2-N}    .  \label{(I.10)}
\end{equation}

The algebra is constructed such that there exists the $q$-analogue of the
exterior derivative $d\equiv C_{ij}dx^i\p^j$ satisfying nilpotency and
Leibniz rule.
In this algebra, one can find the natural $q$-analogue of the Laplacian
$\Delta$ :

\begin{equation}
\Delta=(\p\cdot\p)=C_{ij}\p^i \p^j     ,  \label{(I.11)}
\end{equation}

The $q$-deformed Laplacian of the differential calculus on the N-dimensional
$q$-Euclidian space led us to investigate the corresponding Schr\"odinger
equation, the simplest example of which is the harmonic oscillator. The
action of its Hamiltonian onto the wave function $\big|\Psi\big>$ is defined
by
\begin{equation}
H(\omega)\big|\Psi\big>\, =[-q^{N}(\p\cdot\p)+\omega^2(x\cdot
x)]\big|\Psi\big>\, =\, E\big|\Psi\big>\ ,
\label{(I.12)}
\end{equation}
where we have shifted the normalization to the one introduced in
ref.\cite{[Fio1]} which has the advantage that the factors $q$ in the
corresponding operators are distributed symmetrically w.r.t. the
$*$-conjugation. In this paper we do not write explicitly the result
of the $*$-conjugated sector.  The calculations and proofs given
in the following sections can
be performed completely parallel for the conjugated sector.
In section 5, we discuss the properties
of the system investigated
in this paper with respect to the $*$-conjugation.

\bigskip

As a result of the investigations performed in ref.\cite{[CSW]} we obtained
the solution of the $q$-deformed Schr\"odinger equation for the ground state
of the $q$-deformed harmonic oscillator and the first two excited energy
levels.
The ground state wave function is given by the $q$-exponential function as
\begin{equation}
\big|\Psi_0\big>\,={\rm exp}_{q^2}\bigg[{-\omega xCx\over q^N\mu}\bigg] ,
\label{(I.13)}
\end{equation}
with convention $xCx=C_{ij}x^ix^j$. For the definition of the
$q$-exponential function and some of its
properties see the appendix.
The ground state energy is
\begin{equation}
H(\omega)\big|\Psi_0\big>\,=E_0\big|\Psi_0\big> , \quad{\rm where}\quad
E_0={\omega\mu(1-q^N)\over
(1-q^2)}=\omega Q_N   ,  \label{(I.14)}
\end{equation}
For the first excited level we have the eigenfunction of the vector
representation $\big|\Psi^i_{1}\big>$:
\begin{equation}
H(\omega)\big|\Psi^i_1\big>\,=E_1\big|\Psi^i_1\big> , \with
E_1={\omega\mu(1-q^{N+2})\over q (1-q^2)}
,  \label{(I.15)}
\end{equation}
where
\begin{equation}
\big|\Psi^i_{1}\big>=x^i {\rm exp}_{q^2}
\bigg[{-\omega xCx\over q^{N+1}\mu}\bigg]\ ,
\end{equation}
and for the second excited levels the symmetric tensor
$\big|\Psi_{2,S}\big>$ and singlet representation $\big|\Psi_{2,1}\big>$
with the same energy eigenvalue $E_2$:
\begin{eqnarray}
H(\omega)\big|\Psi^{ij}_{2,S}\big>\,&=&E_2\big|\Psi^{ij}_{2,S}\big> ,
\label{(I.16)}\\
H(\omega)\big|\Psi_{2,1}\big>\,&=&E_2\big|\Psi_{2,1}\big> ,  \with
E_2={\omega\mu(1-q^{N+4})\over
q^2 (1-q^2)}\ . \label{(I.17)}
\end{eqnarray}
The corresponding wave functions are given by
\begin{equation}
\big|\Psi_{2,S}\big> = {\cal P}_S(x\otimes x) {\rm exp}_{q^2}
\bigg[{-\omega xCx\over
q^{N+2}\mu}\bigg]\ ,
\end{equation}
and
\begin{equation}
 \big|\Psi_{2,1}\big> = (xCx + A){\rm exp}_{q^2}
 \bigg[{-\omega xCx\over
 q^{N+2}\mu}\bigg]\ ,
\end{equation}
where $i,j=1,...N$ and $A=-{Q_N q^2\over \omega(1+q^2)}$. All quantities
given here are in the new normalization corresponding to eq.(\ref{(I.12)}).
As for the $q$-numbers we use the conventions:
\begin{equation}
\bra x\ket={q^x-q^{-x}\over q-q^{-1}}   , \label{(I.18)}
\end{equation}
and
\begin{equation}
(x)_{q^2}={1-q^{2x}\over 1-q^2}  . \label{(I.19)}
\end{equation}
These are the results which will be relevant for us in the following.

This paper is organized as follows.  In section 2, we present the main
results of the operator formalism as well as of the $q$-analysis. In section
3, we give the proofs of some theorems and propositions stated in section 2.
In section 4, some further properties of the operator formalism are
analyzed. Section 5 is devoted to discussions and conclusions.

\sectione{The $q$-deformed creation and annihilation operators and general
solution}

\subsection{Operator formalism}

As in ordinary quantum mechanics, to generate all eigenfunctions of the
$q$-deformed Schr\"odinger equation, it is natural to look for the creation
and annihilation operator $a^i$ which satisfies in general

\begin{equation}
H(\omega) a^i=q^k a^i[H(\omega)+C(\omega)]\ ,
\label{(O11)}
\end{equation}
where we introduced a possible $q$-factor $q^k$ ($k$ is a real constant).
Then, $a^i$ maps the $p$th state to the $(p+1)$th state and the constant
$C(\omega)$ gives the energy difference between the two states.  However in
a $q$-deformed system we can easily see that there is not such an operator.
The reason is that the energy difference between the neighbouring states is
not equidistant as we see from the eigenvalues
$E_0$, $E_1$ and $E_2$. Taking this into account, Fiore introduced in
ref.\cite{[Fio1]} operators separately for each state which raise and lower
the energy level, i.e., the $p$-th raising operator $a^\dagger_p$ acts as :
$|p>\rightarrow |p+1>$ (and correspondingly the lowering operator $a_p$) for
the $p$th level.  In this way he obtains all the states together
with an infinite number of raising (and lowering) operators.
 It is a priori not obvious whether one can define at all a
 creation-annihilation operator which produces all states of this
 $q$-deformed system.

The key point to find the creation-annihilation operator of this system is
to allow the quantity $C(\omega)$ to be a function of the Hamiltonian.  One
can easily see that with this generalization the operator $a^i$ still maps
one eigenstate to another eigenstate of different energy level. Using the
analogy with the non-deformed case we look for a creation-annihilation
operator of the form  $(\p^i+x^i\alpha)$ and the above generalization means
that the coefficient $\alpha$ is a function of the Hamiltonian.

In our construction we also have to take into account that the coordinate
function $x^i$ and the derivative $\p^i$ have non-trivial commutation
relations with the Hamiltonian:

\begin{eqnarray}
\p^iH(\omega) &=&H(q\omega)\p^i+\mu\omega^2x^i \ , \label{(2.6.2)}\\
x^iH(\omega)&=&q^{-2}H(q\omega)x^i+q^{N-2}\mu\p^i \ , \label{(2.7.3)}
\end{eqnarray}
With the above described considerations we find the following operators:

\bigskip

\theorem{A}

\noindent{\bf i)}
The creation operator $a^i_-$ and annihilation operator $a^i_+$ are
defined by
\begin{equation}
a^i_{\pm}=q^{-\half}\lambda^{-\half}[q^{N\over2}\p^i+x^i\alpha_\pm(\omega)]\
,
\label{(2.19.4)}
\end{equation}
where
\begin{equation}
\alpha_\pm(\omega)=\half[KH(\omega)\pm\sqrt{K^2[H(\omega)]^2+4\omega^2}]
\with K={(1-q^2)\over q^{N\over2}\mu} \ ,  \label{(2.11.5)}
\end{equation}
and
\begin{equation}
\lambda^{-\half}x^i=q^{-\half}x^i\lambda^{-\half}\and
\lambda^{-\half}\p^i=q^{\half}\p^i\lambda^{-\half}\ .
\label{(2.xx.6)}
\end{equation}

\noindent{\bf ii)}
The commutation relation of the creation and annihilation operators
with the Hamiltonian is
\begin{equation}
H(\omega)a^i_\pm =q^{-1}a^i_\pm [H(\omega)-q^{{N\over2}}\mu
\alpha_\pm(\omega)]\ .  \label{(2.23.7)}
\end{equation}

\bigskip

The proof of this theorem is given in the section 3.1.

\medskip

With these operators defined in Theorem A
we can derive the whole set of states of the corresponding $q$-deformed
Schr\"odinger equation as follows.
First we derive the energy spectrum generated by this operator.  Formula
(\ref{(2.23.7)}) gives the recursion formula of the energy levels from $E_p$
to $E_{p+1}$ when both sides are evaluated on the state $\big|\Psi_p\big>$.
Therefore we
only need to know the eigenvalue of the ground state.
As we have shown in the ref.\cite{[CSW]}, the wavefunction corresponding to
the ground state in the limit $q\rightarrow1$ is given by
$\big|\Psi_0\big>$ in
eq.(\ref{(I.13)}) and it is a candidate of the ground state for the
$q$-deformed case.
We can prove that the operator $a^i_+$ annihilates
$\big|\Psi_0\big>$:

\proposition{A}
\begin{equation}
a^i_+\big|\Psi_0\big>=0     \ .         \label{(2.kk.8)}
\end{equation}

\proof
Using that the ground state energy $E_0=Q_N\omega$
\begin{eqnarray}
\alpha_\pm(\omega)\big|\Psi_0\big>
&=&\half[KE_0\pm\sqrt{K^2[E_0]^2+4\omega^2}]\big|\Psi_0\big>\nonumber\\
&=&\bigg\{ \matrix{q^{-{N\over2}}\omega\big|\Psi_0\big> &\for &\alpha_+& \cr
                            &&&\cr
		-q^{{N\over2}}\omega\big|\Psi_0\big>&\for&\alpha_- &,\cr} \
\label{(2.31.9)}
\end{eqnarray}
where the definition of $K$ is given in eq.(\ref{(2.11.5)}).
Knowing this we act with the annihilation operator $a^i_+$ onto $|\Psi_0>$
and obtain

\begin{eqnarray}
a^i_+\big|\Psi_0\big>\,
&=&q^{-\half}\lambda^{-\half}[q^{N\over2}\p^i
+x^i\alpha_+(\omega)]\big|\Psi_0\big>\nonumber\\
&=&q^{-\half}\lambda^{-\half}\Big[q^{N\over2}q^{-N}\omega
-q^{-{N\over2}}\omega\Big]x^i\big|\Psi_0\big>=0 \ .
\label{(2.32.10)}
\end{eqnarray}
\qed

The excited states are obtained by successively applying the creation
operator $a^i_-$
onto the ground state, the energy eigenvalue of which is defined by
eq.(\ref{(I.14)}).  As we prove in section 3.2, after some calculation we
get the energy eigenvalue of the $p$-th level:

\proposition{B}
\begin{equation}
E_p={\omega\mu\over q^{1-N/2}}\bra{N\over 2}+p\ket \ .   \label{(2.YY.11)}
\end{equation}
\proof Proof is given in section 3.2.

Another result which is obtained in the course of deriving the
energy eigenvalue is the value of the operator $\alpha$ when acting on the
state $|\Psi_p>$:

\begin{equation}
\alpha_\pm(\omega)\big|\Psi_p\big>\,
=\bigg\{\matrix{q^{-p-{N\over2}}\omega\big|\Psi_p\big>&\for &\alpha_+&\cr
&&&\cr
-q^{p+{N\over2}}\omega\big|\Psi_p\big>&\for &\alpha_-& .\cr}
\label{(2.xs.12)}
\end{equation}

Thus as a consequence, we obtain the equation
\begin{eqnarray}
a^i_\pm\big|\Psi_p\big>\,
&=&q^{-\half}\lambda^{-\half}[q^{N\over2}\p^i
+x^i\alpha_\pm(\omega)]\big|\Psi_p\big>\nonumber\\
&=&q^{-\half}\lambda^{-\half}
[q^{N\over2}\p^i\pm x^iq^{\mp(p+{N\over2})}\omega]
\big|\Psi_p\big> \ . \label{(2.33.13)}
\end{eqnarray}
Therefore, when the creation-annihilation operator defined in
eq.(\ref{(2.19.4)}) is acting onto an eigenstate and we evaluate only the
operator $\alpha_\pm(\omega)$ then the resulting expression becomes
level-dependent and coincides with the raising operator
constructed by Fiore.

One of the important relations to characterize the operators
$\alpha_\pm(\omega)$ and $a^i_\pm$ are the following commutation relations

\proposition{C}

\begin{eqnarray}
\alpha_\pm(\omega)a^i_+&=&q^{\pm1}a^i_+\alpha_\pm(\omega)\
,\label{(2.34.14)}\\
\alpha_\pm(\omega)a^i_-&=&q^{\mp1}a^i_-\alpha_\pm(\omega)\
,\label{(2.35.15)}
\end{eqnarray}
and
\begin{equation}
\lambda^{-\half}\alpha_\pm(\omega)=q\alpha_\pm(\omega/q)\lambda^{-\half}\
,\label{(2.xxx.16)}
\end{equation}
the proof of which is given in section 3.3.

Using the above relations we can show that

\theorem{B}

The $q$-antisymmetric product of the creation operator vanishes
\begin{equation}
{\cal P}_A(a^i_-a^j_-)=0\ .
\label{(2.50.17)}
\end{equation}
\proof

\begin{eqnarray}
a^i_-a^j_-
&=&q^{-\half}\lambda^{-\half}[q^{N\over2}\p^ia^j_- +qx^ia^j_-
\alpha_-(\omega)]\nonumber\\
&=&q^{-3/2}\lambda^{-1}[q^{N}\p^i\p^j+q^{N\over2}(\p^ix^j+q^2x^i\p^j)
\alpha_-(\omega)+q^2x^ix^j\alpha_-(\omega)
\alpha_-(\omega)]\ .\label{(2.51.18)}
\end{eqnarray}
Multiplying the projection operator onto both sides and using the defining
relations (\ref{(I.2)}) (\ref{(I.5)}) (\ref{(I.6)}), we get Theorem B.

\qed

Note that the same relation can be proven for the annihilation
 operator, i.e. ${\cal P}_A(a^i_+a^j_+)=0$.

Theorem B means that the creation operator satisfies the same commutation
relation as the $q$-space coordinate function $x^i$. Thus for example
\begin{equation}
(a_-\cdot a_-) a^i_-=a^i_-(a_-\cdot a_-)\ ,
\label{(2.52.19)}
\end{equation}
where $(a_-\cdot a_-)=C_{ij}(a^i_- a^j_-)$.
Consequently any state constructed by successively applying the creation
operator $a_-^i$ onto the ground state $\Psi_0$ is a $q$-symmetric tensor.
Thus we may call the creation-annihilation operator $a_{\pm}^i$ a
$q$-bosonic operator.
Let us state the above results as a theorem.

\theorem{C}

The states of the $p$th level constructed by $p$ creation operators $a^i_-$
\begin{equation}
\big|\Psi_p^{i_1i_2\cdots i_p}\big>\,\equiv a_-^{i_1}a_-^{i_2}\cdots
a_-^{i_p}\big|\Psi_0\big>\ ,
\label{(C1)}
\end{equation}
have the energy eigenvalue $E_p=\omega\mu
q^{{N\over2}-1}\bra{N\over2}+p\ket$ and are $q$-symmetric tensors, i.e.,
$\forall{l}\in \{1,...,p\!-\!1\}$
\begin{equation}
{\cal P}_A{}^{jj'}_{i_li_{l+1}}\big|\Psi_p^{i_1\cdots i_li_{l+1} \cdots
i_p}\big>\,=0\ .
\label{(C2)}
\end{equation}

\proof The energy eigenvalue depends only on the number of creation
operators and thus the wavefunctions defined in (\ref{(C1)}) have the same
value for a fixed level $p$.  The eigenvalue $E_p$ is derived in section
3.2.  The
second part of Theorem C is a direct consequence of Theorem B.
\qed

Thus the number of states of the $p$th level is equal to the number of
states of the non-deformed case, i.e., ${\Big({N+p \ \atop p}\Big)}$.  The
$q$-symmetric tensor can be split into symmetric traceless tensors
corresponding to the irreducible representations of $Fun_q(SO(N))$ :

\begin{equation}
Sym\Big(\underbrace{N\otimes N\otimes\cdots\otimes N}_p\Big)
=S_p\oplus
S_{p-2}\oplus\cdots\oplus \Big\{\matrix{ S_1&{\rm for\  odd}\ p \cr
		& \cr
		S_0&{\rm for\ even}\ p } \Big\} \ .\label{IRRED}
\end{equation}
Correspondingly the wave function in eq.(\ref{(C1)}) is split into
irreducible components.
With this operator method,
Theorem C defines in principle all eigenfunctions.
However it is not straightforward to obtain an expression of the
eigenfunctions as $q$-polynomials in $x^i$.

\subsection{The wave function as a $q$-polynomial in $x$}

In order to obtain an expression of the eigenfunctions in terms of
$q$-polynomials in the coordinate, a simple way is to use the relations of
the $q$-differential calculus with the $q$-analysis \cite{[Exton]}.

The $q$-symmetric traceless $p$th tensor representation $S_p^I$ can be
constructed as follows:
\begin{equation}
S^I_p
\equiv S^{i_1\cdots i_p}_p(x)= S_p{}^{i_1\cdots i_p}_{j_1\cdots
j_p}x^{j_1}\cdots x^{j_p} \ ,  \label{(2.53.20)}
\end{equation}
where
\begin{eqnarray}
{\cal P}_A{}^{kl}_{i_ji_{j+1}}S_p^{i_1\cdots i_ji_{j+1}\cdots
i_p}(x)&=&0 \ , \label{(2.35.21)}\\
C_{i_ji_{j+1}}S_p^{i_1\cdots i_ji_{j+1}\cdots i_p}(x)&=&0 \ ,
 \label{(2.36.22)}
\end{eqnarray}
for $j=1,...,p-1$.

Since the tensor structure is defined by these $q$-symmetric tensors, the
wave functions can be written by the product of $q$-symmetric tensor and
function of $x^2$. Thus to solve the Schr\"odinger equation (\ref{(I.12)}),
we take the ansatz:
\begin{equation}
\big|\Psi\big>\,=S^I_pf(x^2)\ .
\label{(P1)}
\end{equation}
The problem is to fix the function $f(x^2)$.
For this end, we need the following formulae:

\begin{eqnarray}
\Delta S^I_p
&=&\mu(p)_{q^2}(C_{ij}S^{Ii}\p^j)+q^{2p}S^I_p\Delta \ , \label{(a.23)}\\
\p^if(x^2)&=&\mu x^i D_{x^2}f(x^2) \ , \label{(b.24)}\\
\Delta f(x^2)&=&[q^N\mu^2 x^2 D^2_{x^2}+\mu
Q_ND_{x^2}]f(x^2) \ , \label{(c.25)}\\
\Delta S^I_p
f(x^2)&=&\mu^2S_p^I[q^{N+2p}x^2D^2_{x^2}
+({N\over2}+p)_{q^2}D_{x^2}]f(x^2) \ .\label{(dd.26)}
\end{eqnarray}

Computing the action of the Hamiltonian onto the wave function (\ref{(P1)}),
the Schr\"odinger equation becomes:
\begin{eqnarray}
(-q^N\Delta+\omega^2 x^2) S^I_p
f(x^2)&=&\mu^2S_p^I[-q^{2N+2p}x^2D^2_{x^2}
-q^N({N\over2}+p)_{q^2}D_{x^2}+{\omega^2\over\mu^2}x^2]f(x^2)\nonumber\\
   &=&E S_p^If(x^2) \ , \label{(e.27)}
\end{eqnarray}
where the definition of $D_{x^2}$ is
\begin{equation}
D_{x^2}f(x^2)={f(q^2 x^2)-f(x^2)\over x^2(q^2-1)} \ .   \label{(d.28)}
\end{equation}

{}From eq.(\ref{(e.27)}) we get the following $q$-difference equation for
$f(x^2)$
\begin{equation}
F(D_{x^2})f(x^2)
=[-q^{2N+2p}x^2D^2_{x^2}-q^N({N\over2}+p)_{q^2}D_{x^2}
+{\omega^2\over\mu^2}x^2-{E\over\mu^2}]f(x^2)=0\ .\label{(DE)}
\end{equation}
To solve this equation we take an ansatz with the $q$-exponential function
as:
\begin{equation}
f(x^2)=\sum b_s\qexp(q^{-2s}\alpha)\where \alpha=q^{-N-p}\omega \ .
\label{(DE1)}
\end{equation}
The definition and properties of the $q$-exponential function
$\qexp(\alpha)$ are collected in the appendix.
A lengthy but straightforward calculation yields
\begin{eqnarray}
F(D_{x^2})f(x^2)
&=&{q^{{N\over2}-1}\omega\over\mu}
\sum\Big[-b_{s+1}q^{{N\over2}+p}[\![2s+2]\!]\nonumber\\
&\quad&\qquad\qquad+b_s\Big([\![{N\over2}+p+2s]\!]
-{E\over q^{{N\over2}-1}\mu\omega}\Big)\Big]\qexp(q^{-2s}\alpha)\ .
\label{(DF)}
\end{eqnarray}

We look for the solution which has a finite number of terms in the expansion
 eq.(\ref{(DE1)}). In such a case the argument of the $q$-exponential
 functions in
eq.(\ref{(DE1)}) can be shifted to the $q$-exponential of the largest $s$ by
using the relations given in the appendix. Then such a function $f(x^2)$
becomes a polynomial of $x^2$ multiplied with the $q$-exponential which
has a smooth finite limit under $q\rightarrow1$.
On the other hand since the $q$-exponentials with different arguments
generate different powers in $x^2$, they are independent and cannot cancel
each other.  Thus solving the equation, we require that for each term in the
series eq.(\ref{(DF)}) the sum of the coefficients of different exponential
functions separately vanishes.

First we see that in the term for $\qexp(\alpha)$, i.e. for $s=-1$ in
eq.(\ref{(DF)}), the coefficient
of $b_0$  is zero. Therefore we can consistently set $b_s=0$ for $s<0$ and
require that the first nonzero term starts with the $b_0$ term.

Now we require that the series has only a finite number of terms.  To
satisfy this, the second term under the sum, i.e., the coefficient of
$\qexp(q^{-2s}\alpha)$ containing the factor $b_s$
must vanish for a certain $s$.  Calling this largest integer $s$ as $r$,
this requirement defines the energy eigenvalue $E$ as

\begin{equation}
E=q^{{N\over 2}-1}\mu\omega[\![{N\over 2}+p+2r]\!] \ .
\label{(2.54.29)}
\end{equation}

With this eigenvalue, we can set the $b_s=0$ for $s>r$ and can solve the
equation (\ref{(DE)}).
{}From the condition that the sum of all terms with the same argument in the
exponential function vanishes we get the recursion formula

\begin{equation}
b_s=
b_{s+1}{-q^{{N\over2}+p}[\![2s+2]\!]\over[\![{N\over2}+p+2r]\!]
-[\![{N\over2}+p+2s]\!]} \ .
\label{(2.56.30)}
\end{equation}
Therefore we obtain

\begin{equation}
b_s= b_r\prod_{s\le t\le r-1} {q^{{N\over2}+p}[\![2t+2]\!]\over
[\![{N\over2}+p+2t]\!]-[\![{N\over2}+p+2r]\!]} \ .
\label{(coeff)}
\end{equation}
The $b_r$ simply shows the freedom of the overall normalization and thus the
wave
functions are now defined in terms of the $q$-exponential functions and the
$q$-polynomials of the coordinate functions $x^i$ with a finite number of
terms and with the $b_s$ given above as

\begin{equation}
\big|\Psi_{p,r}^I\big>\,=S_p^I\sum_{s=0}^{r} b_s\qexp(q^{-2s}\alpha)\where
\alpha=q^{-N-p}\omega \ .\label{(DE2)}
\end{equation}

The energy eigenvalue of $|\Psi_{p,r}>$ is given by $E$
in eq.(\ref{(2.54.29)}) and it coincides with the
eigenvalue defined by using the creation-annihilation operator in the
previous section. We can also confirm that there is a one-to-one
correspondence between the wave function given in eq.(\ref{(C1)}) and the
one in eq.(\ref{(DE2)}):

The wave function derived in eq.(\ref{(DE2)}) shows that for each $p$th rank
tensor we have an infinite tower of eigenfunctions labeled by the integer
$r$ with the eigenvalue $E_{p+2r}=q^{{N\over 2}-1}\mu\omega[\![{N\over
2}+p+2r]\!]$.  This means that for the fixed eigenvalue $E_{p'}$ there is
one  eigenfunction of $p$th rank tensor for each $p$ which satisfies
$p+2r=p'$ with a positive integer $r$. This is the result given in
eq.(\ref{IRRED}).

This completes the $q$-analytic construction of the eigenfunctions which
gives the $q$-polynomial representation of the wave function corresponding
to the irreducible representations of the $Fun_q(SO(N))$.

\sectione{Proofs of Theorem A, Proposition B and Proposition C}

In this section we give the proof of Theorem A, the energy eigenvalue
$E_p$ given in Proposition B, as well as of Proposition C.

\bigskip

\subsection{Proof of Theorem A}

Since the commutation relations of $H(\omega)$ with $x$ and the one with
$\p$ generate a discrepancy in the factors $q$ as one can see from
eqs.(\ref{(2.6.2)}) and (\ref{(2.7.3)}), we carefully have to choose the
operator $\alpha$.  It turns out that when we put the operator $\alpha$ at
the right hand side of $x^i$, we have to consider $\alpha$ as a function of
$H(\omega/q)$.  Taking this into account we consider the following operator:
\begin{equation}
A_{\pm}(\omega)=[q^{N\over2}\p^i+ x^i\alpha_\pm(\omega/q)] , \label{(EEE)}
\end{equation}
where $\alpha(\omega/q)$ is a function of the Hamiltonian $H(\omega/q)$ and
thus $\alpha(\omega/q)$ does not commute with $H(\omega)$ but with
$H(\omega/q)$.

The commutation relation of the operators $A_{\pm}$ in eq.(\ref{(EEE)}) with
the Hamiltonian can be computed by using
eqs.(\ref{(2.6.2)})-(\ref{(2.7.3)}):

\begin{eqnarray}
H(\omega)A_{\pm}(\omega)
&=&q^{N\over2}\p^i[H(\omega/q)-q^{N\over2}\mu\alpha( \omega/q)] \nonumber\\
&\quad&\qquad\quad
+x^i[q^2H(\omega/q)\alpha(\omega/q)-q^{N\over2}\mu\omega^2q^{-2}]\ .
\label{(2.12)}
\end{eqnarray}
We require that the r.h.s. is also proportional to the operator
$A_\pm(\omega)$.  Thus $\alpha(\omega/q)$ is defined by the condition
\begin{equation}
[H(\omega/q)-q^{N\over2}\mu\alpha(\omega/q)]\alpha(\omega/q)
=[q^2H(\omega/q)\alpha(\omega/q)-q^{N\over2}\mu\omega^2q^{-2}] \ .
\label{(2.13)}
\end{equation}
Since the operator $\alpha(\omega/q)$ commutes with the $H(\omega/q)$, the
above equation leads to the simple quadratic equation for $\alpha(\omega/q)$
\begin{equation}
[\alpha(\omega/q)]^2-{(1-q^2)\over
q^{N\over2}\mu}H(\omega/q)\alpha(\omega/q)-{\omega^2\over q^2}=0\ ,
\label{(2.14)}
\end{equation}
the solution of which is given by eq.(\ref{(2.11.5)}). Then from
eq.(\ref{(2.12)}) we get

\begin{equation}
H(\omega)A_{\pm}(\omega)
=A_\pm(\omega)[H(\omega/q)-q^{N\over2}\mu\alpha(\omega/q)]\ .
\label{(3.1)}
\end{equation}

This is not a commutation relation yet since
the argument of the Hamiltonian is shifted. In order to obtain
eq.(\ref{(2.23.7)}) we still have to improve our operator such that the
argument of the Hamiltonian remains also unchanged. This can be achieved by
introducing the shift operator $\lambda$, the action of which is defined as

\begin{equation}
\lambda x =qx\lambda  \ , \label{(2.15)}
\end{equation}

\begin{equation}
\lambda \p =q^{-1}\p\lambda \ .  \label{(2.16)}
\end{equation}
Consequently we get
\begin{equation}
\lambda^{-\half} H(\omega)=qH(\omega/q)\lambda^{-\half} \ ,\label{(2.17)}
\end{equation}
and
\begin{equation}
\lambda^{-\half}\alpha_\pm(\omega)=q\alpha_\pm(\omega/q)\lambda^{-\half}\ .
\label{(2.18)}
\end{equation}
Note that the shift operator $\lambda$ relates to the algebra element
$\Lambda$ introduced in ref.\cite{[OZ]} by
\begin{equation}
\lambda^2=\Lambda=1 +(q^2-1)(x\cdot\p) +{(q^2-1)^2\over \mu^2
q^{N-2}}(x\cdot
x)\Delta  \ .   \label{(3.2)}
\end{equation}

Thus we define improved operators by including the operator $\lambda$ as

\begin{equation}
a_{\pm}=[q^{N\over2}\p^i+x^i\alpha_\pm(q^{-1}\omega)]\lambda^{-\half}\ .
\label{(2.19)}
\end{equation}

Then using eqs.(\ref{(2.6.2)})-(\ref{(2.7.3)}) we get

\begin{equation}
H(\omega)a_\pm=q^{-1}a_\pm H(\omega)-q^{{N\over2}-1}\mu
a_\pm\alpha_\pm(\omega)\ . \label{(2.23)}
\end{equation}
\qed

\bigskip

\subsection{Proof of the general form of the energy eigenvalue $E_p$}

Eq.(\ref{(2.23.7)}) provides a recursion
formula for the energy eigenvalues and
the proof is given by induction:
$E_0$ is given by acting with the Hamiltonian on the ground state $\Psi_0$,
we get $E_0={\omega\mu\over q^{1-N/2}}\bra {N\over2}\ket$.
Acting with $H(\omega)a^i_-$ onto some eigenfunction $|\Psi_p>$ we obtain
according to eq.(\ref{(2.23.7)})
\begin{equation}
H(\omega)a^i_-\big|\Psi_p\big>\, =
a^i_-q^{-1}[E_p-\mu\alpha_-(\omega)q^{N\over
2}]\big|\Psi_p\big>\ . \label{(MM)}
\end{equation}

Suppose
\begin{equation}
E_{p}=q^{{N\over 2}-1}\omega\mu \bra{{N\over 2}+p}\ket\ ,\label{(3.3)}
\end{equation}
then
\begin{eqnarray}
\alpha_\pm(\omega)\big|\Psi_p\big>\,
&=&\half[KE_p\pm\sqrt{K^2[E_p]^2+4\omega^2}]\big|\Psi_p\big>\nonumber\\
&=&\Big\{\matrix{ q^{-p-{N\over2}}\omega\big|\Psi_p\big>&\for &\alpha_+&\cr
		&&&\cr
                -q^{p+{N\over2}}\omega\big|\Psi_p\big>&\for &\alpha_-& .\cr}
\label{(3.4)}
\end{eqnarray}

Substituting this into eq.(\ref{(MM)}) we obtain

\begin{eqnarray}
H(\omega)a^i_-\big|\Psi_p\big>\,
&=&a^i_- q^{-1}\bigg[E_p + q^{N+p} \mu\omega
\bigg]\big|\Psi_p\big>\nonumber\\
&=&a^i_- \omega\mu {1\over q^{1-{N\over 2}}}\bra{{N\over
2}+p+1}\ket\big|\Psi_p\big>
\nonumber\\
&=&E_{p+1}a^i_- \big|\Psi_p\big>\ , \label{(3.5)}
\end{eqnarray}
thus
\begin{equation}
E_{p+1}=\omega\mu {1\over q^{1-{N\over 2}}}\bra{{N\over 2}+p+1}\ket\
.\label{(3.6)}
\end{equation}
Therefore the energy eigenvalue of the $p$-th level is given by
\begin{equation}
E_p={\omega\mu(1-q^{N+2p})\over q^{N/2+p}(1-q^2)}={\omega\mu\over
q}\bra{N\over 2}+p\ket \ .  \label{(YY)}
\end{equation}
which is the result stated in Proposition B.

\qed

\subsection{ Proof of Proposition C}

In Proposition C the
commutation relation of $a_{\pm}$ and $\alpha_{\pm}$
is stated, the proof of which is as follows:

\proof

\begin{eqnarray}
\alpha_\pm(\omega)a_+
&=&\half[KH(\omega)\pm\sqrt{K^2[H(\omega)]^2+4\omega^2}]a_+\nonumber\\
&=&a_+\half\Big[K[q^{-1} H(\omega)-q^{{N\over2}-1}\mu
\alpha_+(\omega)]\nonumber\\
&\quad& \pm\sqrt{K^2[q^{-1} H(\omega)-q^{{N\over2}-1}\mu
\alpha_+(\omega)]^2+4\omega^2}\Big] \ .\label{(3.8)}
\end{eqnarray}
The expression under the square root can be rewritten in a
more convenient way as
\begin{eqnarray}
K^2\,[\,q^{-1} H(\omega)&-&q^{{N\over2}-1}\mu
\alpha_+(\omega)\,]\,^2+4\omega^2\nonumber\\
&=&q^{-2}K^2 H^2+[-2q^{-2}(1-q^2)
KH+q^{-2}(1-q^2)^2\alpha_+-4\alpha_-]\alpha_+\nonumber\\
&=&q^{-2}\Big[K H-(1+q^2)\alpha_+\Big]^2 \ , \label{(3.10)}
\end{eqnarray}
where we have used that $\omega^2=-\alpha_+\alpha_-$.
Thus
\begin{equation}
\sqrt{K^2[q^{-1} H(\omega)-q^{{N\over2}-1}\mu \alpha_+(\omega)]^2+4\omega^2}
=q^{-1}[(1+q^2)\alpha_+-KH]\ ,\label{(3.11)}
\end{equation}
where, in order to determine the overall sign we have required that for the
case $q=1$ (no deformation) the value of the square root be $\sqrt{\
}=+\omega$.
Correspondingly for the case of $\alpha_-$ we obtain
\begin{equation}
\sqrt{K^2[q^{-1} H(\omega)-q^{{N\over2}-1}\mu \alpha_-(\omega)]^2+4\omega^2}
=q^{-1}[(1+q^2)\alpha_--KH]\ ,\label{(3.12)}
\end{equation}
where we again require that $\sqrt{\ }=+\omega$ for $q=1$.
Therefore we obtain the following commutation relations
\begin{eqnarray}
\alpha_\pm(\omega)a_+
&=&a_+\half\Big[K[q^{-1} H-q^{{N\over2}-1}\mu \alpha_+]\pm
q^{-1}[(1+q^2)\alpha_+-KH]\Big]\nonumber\\
&=&\bigg\{\matrix{a_+\half\Big[-Kq^{{N\over2}-1}\mu \alpha_++
q^{-1}(1+q^2)\alpha_+\Big]&\cr
  q^{-1}a_+\half\Big[2K H-[q^{{N\over2}}\mu K + (1+q^2)]\alpha_+\Big]
 &\cr}\bigg\}=q^{\pm 1}a_+\alpha_{\pm} \ , \label{(3.13)}
\end{eqnarray}

\begin{eqnarray}
\alpha_\pm(\omega)a_-
&=&a_-\half\Big[K[q^{-1} H-q^{{N\over2}-1}\mu \alpha_-]\pm
q^{-1}[KH-(1+q^2)\alpha_-]\Big]\nonumber\\
&=&\bigg\{\matrix{a_-\half\Big[-Kq^{{N\over2}-1}\mu \alpha_-+
q^{-1}(1+q^2)\alpha_-\Big]&\cr
  q^{-1}a_+\half\Big[2K H-[q^{{N\over2}}\mu K + (1+q^2)]\alpha_-\Big]
  &\cr}\bigg\}=q^{\pm 1}a_-\alpha_{\mp} \ . \label{(3.14)}
\end{eqnarray}
\qed

\sectione{Operator relations}

It is interesting to ask how far we can make the analogy of the
operator algebra using the creation-annihilation operator. We investigate
here further properties of the creation-annihilation operator given in
eq.(\ref{(2.19.4)}) and give some results concerning the operator formalism
of the $q$-deformed Schr\"odinger
equation (\ref{(I.12)}).

First we give an alternative representation of the creation-annihilation
operator.
When we defined the creation-annihilation operator we have taken the ansatz
(\ref{(EEE)}) where the operator $\alpha$ is on the right hand side of the
coordinate $x^i$.  Actually we can find the solution
with the ansatz where the operator $\alpha'$ is on the left hand side of
the coordinate $x^i$: $b^i=\lambda^{-\half}(\p^i+\alpha' x^i)$. The
derivation of the operator $\alpha'$ is analogous to the one given in
section 3.1. The result is

\theorem{D}
The creation-annihilation operator can be also represented by

\begin{equation}
b_\pm=\lambda^{-\half}[q^{N\over2}\p^i+\alpha_\pm(q\omega)x^i]\ ,
\label{ThD0}
\end{equation}
where the operator $\alpha(\omega)$ is given by eq.(\ref{(2.11.5)}). These
two representations satisfy the identity
\begin{equation}
{1\over\sqrt{K^2[H(\omega)]^2+4\omega^2}}b_\pm
=q^{3\over2} a_\pm{1\over\sqrt{K^2[H(\omega)]^2+4\omega^2}}\ , \label{ThD}
\end{equation}

\proof
One can derive $b_\pm$ directly as performed for $a_\pm$ and
show that both operators eq.(\ref{(2.19.4)}) and
eq.(\ref{ThD0}) give the same energy eigenvalues of the states.
Here, instead of repeating these calculations as performed in section 3 we
give the proof of the
relation (\ref{ThD}), which also proves the above statements.

First we prove
the commutator of $\alpha$ with $x$, which is given in a convenient
form by including the shift operator $\lambda$ as
\begin{eqnarray}
\alpha_{\pm}(x^i\lambda^{-{1\over 2}})-{1\over q}(x^i\lambda^{-{1\over
2}})\alpha_{\pm}&=&\pm (q^2-1)a_{\pm}{\alpha_{\pm}(\omega)\over \sqrt{K^2
[H(\omega)]^2+4\omega^2}}\nonumber\\
&=&\pm (q^2-1)a_{\pm}{\alpha_{\pm}(\omega)\over
(\alpha_+(\omega)-\alpha_-(\omega))}\ . \label{ThD1}
\end{eqnarray}

Substituting
\begin{equation}
a_\pm=[q^{N\over2}\p^i+x^i\alpha_\pm(q^{-1}\omega)]\lambda^{-\half}\ ,
\end{equation}
into Proposition C we get

\begin{eqnarray}
\alpha_\pm(\omega)[q^{N\over2}\p^i
+x^i\alpha_+(q^{-1}\omega)]\lambda^{-\half}
&=&q^{\pm1}[q^{N\over2}\p^i
+x^i\alpha_+(q^{-1}\omega)]\lambda^{-\half}\alpha_\pm(\omega)\ ,\\
\alpha_\pm(\omega)[q^{N\over2}\p^i
+x^i\alpha_-(q^{-1}\omega)]\lambda^{-\half}
&=&q^{\mp1}[q^{N\over2}\p^i
+x^i\alpha_-(q^{-1}\omega)]\lambda^{-\half}\alpha_\pm(\omega)\ .
\end{eqnarray}

Subtracting the second equation from the first yields
\begin{equation}
\alpha_\pm(\omega)x^i\lambda^{-\half}[\alpha_+(\omega)-\alpha_-(\omega)]
=\pm q(q-q^{-1})a_\pm\alpha_\pm(\omega)+q^{-1}x^i
\lambda^{-\half}[\alpha_+(\omega)-\alpha_-(\omega)]\alpha_\pm(\omega)\ .
\label{ThD3}
\end{equation}

Dividing by the factor $(\alpha_+-\alpha_-)$
we get the relation(\ref{ThD1})

With this results we obtain the relation between the $a_\pm$ and $b_\pm$ as
\begin{eqnarray}
b_\pm
&=&q^\half[a_\pm \pm q(q-q^{-1})a_\pm{\alpha_\pm(\omega)
\over[\alpha_+(\omega)-\alpha_-(\omega)]}]
\nonumber\\
&=&q^{3\over2} a_\pm  {q^{\pm1}\alpha_+(\omega)
-q^{\mp1}\alpha_-(\omega)
\over[\alpha_+(\omega)-\alpha_-(\omega)]}\ .
\label{ThD4}
\end{eqnarray}
Thus we get
\begin{equation}
b_\pm
=q^{3\over2}  [\alpha_+(\omega)-\alpha_-(\omega)]a_\pm{1\over
[\alpha_+(\omega)-\alpha_-(\omega)]}\ .
\label{ThD5}
\end{equation}
Dividing by $(\alpha_+-\alpha_-)$ from the left
we get the equation (\ref{ThD}) of Theorem D.

\qed

The operator $b^i_\pm$ is important when we investigate the transformation
rule of the creation-annihilation operator under the $*$-conjugation.
This problem is beyond the scope of this paper and will be discussed
elsewhere.
We show in the rest of this section how the Hamiltonian can be represented
in terms of the  creation-annihilation operator defined in Theorem A.

Analogously to the non-deformed case the operators
$(a_+\cdot a_-)$ and $(a_-\cdot a_+)$ are singlet representations of the
quantum group.  Thus we expect that they can be given as functions of the
Hamiltonian.
A straightforward computation leads to the following operator identity
\begin{eqnarray}
(a_+\cdot a_-)
&=&q^{-{3\over2}}\lambda^{-1}\widehat{B}[-H(\omega)+Q_Nq^{N\over2}\alpha_-
(\omega)]\ ,\label{(3.15)}\\
(a_-\cdot a_+)
&=&q^{-{3\over2}}\lambda^{-1}\widehat{B}[-H(\omega)+Q_Nq^{N\over2}
\alpha_+(\omega)]\ ,\label{(3.16)}
\end{eqnarray}
and the linear combination of the above expressions gives the following
simple relation
\begin{equation}
(a_+\cdot a_-)+(a_-\cdot a_+)
=-q^{-{3\over2}}(1+q^N)\lambda^{-1}\widehat{B}H(\omega)\ , \label{(d.1)}
\end{equation}
where
\begin{equation}
\widehat{B}=1+{q^2-1\over\mu}(x\cdot\p)\ .
\label{(3.17)}
\end{equation}
The extra operator $\lambda^{-1}\widehat{B}$ commutes with the Hamiltonian:

\begin{equation}
[\lambda^{-1}\widehat{B},H(\omega)]=0\ .
\label{(3.18)}
\end{equation}
Actually the eigenfunctions also form a diagonal basis with respect to this
operator, i.e. when acting with the operator
$\lambda^{-1}\hat B$ onto the wave function eq.(\ref{(P1)}) we obtain
\begin{equation}
\lambda^{-1}\hat B\big|\Psi_p\big>\,={\cal N}_p\big|\Psi_p\big> \
,\label{(3.20)}
\end{equation}
where ${\cal N}_p$ is a certain $q$-number. However, this eigenvalue ${\cal
N}_p$ is not independent of the states.  By using the explicit form of the
wave function $\Psi_{p+2r}=S_pf_r(x^2)$ derived in section 2 we can
determine the eigenvalue ${\cal N}_p$.
It is given by
\begin{equation}
\lambda^{-1}\hat B\big|\Psi_{p+2r}\big>\,={q\over \mu q^{N\over
2}}(q^{-{N\over
2}-p+1}+q^{{N\over 2}+p-1})\big|\Psi_{p+2r}\big> \ .\label{EigenValueOfNp}
\end{equation}
{}From this we see that the eigenvalue of $\lambda^{-1}\hat B$ depends only on
the tensor structure defined by $p$ and is nonzero. To get the Hamiltonian,
 we have to divide out this operator.

\bigskip

Finally we give the commutation relation of $a_+^i$ and
$a_-^j$:
\begin{eqnarray}
{\cal P}_A(a_+^ia_-^j)&=&\lambda^{-1}q^{N-3\over 2}{\cal
P}_A(\p^ix^j\alpha_-(\omega)+x^i\p^j\alpha_+(\omega)) \nonumber\\
&=&\lambda^{-1}q^{N-3\over 2}{\cal
P}_A(x^i\p^j)[\alpha_+(\omega)-q^2\alpha_-(\omega)] \ . \label{(3.22)}
\end{eqnarray}
The operator ${\cal P}_A(x^i\p^j)$ appearing in the r.h.s. is proportional
to the angular momentum in the limit $q\rightarrow1$.  (Some properties of
the operator ${\cal P}_A(x^i\p^j)$ have been discussed in
ref.\cite{[Fio1]}.)
This means that the commutator of the creation and annihilation operator
contains the angular momentum.

\sectione{Discussion and conclusion}

In this paper we have shown two different methods to construct the
solution of the $q$-deformed Schr\"odinger equation with $Fun_q(SO(N))$
symmetry.  It is
proven that a creation-annihilation operator exists which
generate all excitation levels. We also gave the explicit solution in terms
of the $q$-polynomial and $q$-exponential functions by solving the
associated $q$-difference equation.

Concerning the Hamiltonian it is not completely straightforward to express
it in
terms of the creation-annihilation operators as we see from the result of
eq.(\ref{(d.1)}).
One way to investigate such a property is to
 take the operators $a^i_{\pm}$ as the fundamental quantities
of the system, and consider the 'improved' Hamiltonian which is directly
proportional to $((a_+\cdot a_-)+(a_-\cdot a_+))$ on operator level. For
this one may still consider the rescaling of the creation-annihilation
operator by the function of the Hamiltonian as is suggested by
 the theorem D.  From eq.(\ref{ThD}), we see that when we define the
 creation-annihilation operator with the factor
 ${1\over\sqrt{K^2[H(\omega)]^2+4\omega^2}}$ appropriately, the relation
 between the improved $a^i_\pm$ and $b^i_\pm$ is simplified. Such an
 improved   creation-annihilation operator also seems to simplify the
 $*$-conjugation of the operators. This problem is still under
 investigation.

Concerning the conjugation property, there is the following problem which
has to be solved.
Under the simple
$*$-conjugation which is defined by \cite{[OZ]}
\begin{eqnarray}
 x^{i*}&=&x_i \ , \label{Dis0}\\
 \p^{i*}&=&-q^{-N}\bar\p_i\ , \label{Dis1}
\end{eqnarray}
the Hamiltonian is transformed as
\begin{equation}
H^*=-q^{-N}(\bar\p\cdot\bar\p)+\omega^2(x\cdot x) \ ,  \label{Dis2}
\end{equation}
If we apply the reality condition proposed in ref.\cite{[OZ]},
the above equation gives
\begin{equation}
H^*=-q^{-2}{1\over \Lambda}(\p\cdot\p)+\omega^2(x\cdot x)\ ,
\label{Dis2a}
\end{equation}
Thus under the $*$-conjugation with applying the reality condition proposed
in ref.\cite{[OZ]}, the Hamiltonian is not an hermitian operator.

On the other hand considering the various possible Laplacians and its
operations on the $q$-exponential function, we find that
 $(\p\cdot\p)$ and
$(\bar\p\cdot\bar\p)$ have simple relations which can be identified with the
eigenfunction equation.  We listed the four types of Laplacians which
may be considered as the eigenfunction equation in the appendix. These four
Laplacians are all non-hermitian under the $*$-conjugation with the reality
condition given in ref.\cite{[OZ]}. With our present
knowledge, it is not possible to construct the eigenfunctions for a
Laplacian which is hermitian under the conjugation discussed above.

Recall that the hermiticity condition is related with the definition of
the norm. On the other hand the definition of the norm is also related
with the definition of the integration on the quantum space.
In principle the integration can be defined algebraically by requiring the
property of the partial integration.\cite{[HW],[Fio1]}. However, this does
not fix the norm and
we may still consider the possibility of modifying the norm and the
definition of the hermitian conjugation.
One possibility is discussed in ref.\cite{[Fio1]}.  We shall report
elsewhere on this problem, including the integration on the quantum
space.

Finally we would like to remark
that this formulation of the N-dimensional harmonic oscillator on
quantum space
may be applied to the formulation of
the $q$-deformed string theory. See ref.\cite{[Kulish]} and references
therein.

\bigskip

\noindent{\bf Acknowledgement}

One of the authors (U.C.) would like to thank the Tohoku Kaihatsu Kinen
foundation for financial support.

\startappA

\noindent{ \large \bf Appendix}

In this appendix we collected some of the basic properties of the
$q$-exponential functions.  In order to simplify the expressions we
introduce the following abbreviation $\qexp(\alpha)$. Note that for the
summary given in section 1 we followed the conventions of our previous paper
\cite{[CSW]}.
The new notations and their relations to the previous ones are:

The definition of the $q$-exponential function is
\begin{eqnarray}
 \qexp(\alpha) &=&{\rm exp}_{q^2}
\bigg[{-\alpha xCx\over \mu}\bigg]
=\sum {1\over(n)_{q^2}!}(-{\alpha(x\cdot
 x)\over\mu})^n \ , \label{(A.1)}\\
 \bqexp(\alpha) &=&{\rm exp}_{q^{-2}}
\bigg[{-\alpha xCx\over \mu}\bigg]
=\sum {1\over(n)_{q^{-2}}!}(-{\alpha(x\cdot
 x)\over\bar\mu})^n \ , \label{(A.2)}
\end{eqnarray}
where
\begin{equation}
(n)_x={1-x^n\over1-x}, \and \bar\mu=q^{N-2}\mu\ .   \label{(A.3)}
\end{equation}

It is straightforward to derive the following identities

\begin{eqnarray}
\qexp(\alpha)-\qexp(q^2\alpha) &=-(1-q^2){\alpha(x\cdot
x)\over\mu}\qexp(\alpha)\ ,  \label{(A.4)}\\
 \bqexp(\alpha)-\bqexp(q^{-2}\alpha) &=-(1-q^{-2}){\alpha(x\cdot
 x)\over\bar\mu}\bqexp(\alpha)\ .
 \label{(A.5)}
\end{eqnarray}

Using these identities we can easily compute the action of the
derivative $D_{x^2}$ onto the $q$-exponential function. Thus for
$D_{x^2}f(x^2)$ with $f(x^2)$ given in eq.(\ref{(DE1)}) we obtain

\begin{equation}
D_{x^2}\sum b_s\qexp(q^{-2s}\alpha)=-{\alpha\over \mu}\sum
b_s q^{-2s} \qexp(q^{-2s}\alpha)\ .
\label{(a.i)}
\end{equation}

For the discussion on the $*$-conjugation properties it is also necessary to
consider the action of the derivative $\p^i$ and of $\bar{\p}^i$ onto the
$q$-exponential function where the $*$-conjugation is defined as
eqs.(\ref{Dis0}) and (\ref{Dis1}).
The definition of the action of $\p^i$ has been given in eq.(\ref{(I.5)}).
For the derivative $\bar{\p}^i$ we define
\begin{equation}
\bar{\p}^ix^j=C^{ij} +q^{-1}\hat R^{ij}_{kl}x^k\bar{\p}^l\  . \label{(ai.5)}
\end{equation}
Eq.(\ref{(I.5)}) and eq.(\ref{(ai.5)}) are conjugate to each other
under the $*$-operation given in eqs.(\ref{Dis0})
and (\ref{Dis1}).
Then the action of the derivatives onto $\qexp(\alpha)$ yield
\begin{eqnarray}
\p^i \qexp(\alpha) &=&-\alpha x^i \qexp(\alpha)+GT \ , \label{(A.6)}\\
\bar\p^i \qexp(\alpha) &=&-\alpha x^i \qexp(q^{-2}\alpha)+GT \ ,
\label{(A.7)}\\
\p^i \bqexp(\alpha) &=&-\alpha x^i \bqexp(q^2\alpha)+GT\ , \label{(A.8)}\\
\bar\p^i \bqexp(\alpha) &=&-\alpha x^i\bqexp(\alpha)+GT\ . \label{(A.9)}
\end{eqnarray}
The abbreviation GT simply means "go-through term"; in the case of eq.(A.6)
for example $GT=\qexp(q^2 \alpha)\p^i$.
\begin{eqnarray}
(\p\cdot\p)\qexp(\alpha)&=&-\alpha Q_N\qexp(\alpha)+q^N\alpha^2(x\cdot
x)\qexp(\alpha)+GT\ , \label{(A.10)}\\
(\bar\p\cdot\p)\qexp(\alpha)&=&-\alpha
Q_N\qexp(\alpha)+q^{-N}\alpha^2(x\cdot
x)\qexp(q^{-2}\alpha)+GT\ , \label{(A.11)}\\
(\p\cdot\bar\p)\qexp(\alpha)&=&-\alpha
Q_N\qexp(q^{-2}\alpha)+q^{N-2}\alpha^2(x\cdot
x)\qexp(q^{-2}\alpha)+GT\ , \label{(A.12)}\\
(\bar\p\cdot\bar\p)\qexp(\alpha)&=&-\alpha
Q_N\qexp(q^{-2}\alpha)+q^{-N-2}\alpha^2(x\cdot
x)\qexp(q^{-4}\alpha)+GT\ , \label{(A.13)}\\
(\p\cdot\p)\bqexp(\alpha)&=&-\alpha
Q_N\bqexp(q^2\alpha)+q^{N+2}\alpha^2(x\cdot
x)\bqexp(q^4\alpha)+GT\ , \label{(A.14)}\\
(\bar\p\cdot\p)\bqexp(\alpha)&=&-\alpha
Q_N\bqexp(q^2\alpha)+q^{2-N}\alpha^2(x\cdot
x)\bqexp(q^2\alpha)+GT\ , \label{(A.15)}\\
(\p\cdot\bar\p)\bqexp(\alpha)&=&-\alpha Q_N\bqexp(\alpha)+q^N\alpha^2(x\cdot
x)\bqexp(q^{2}\alpha)+GT\ , \label{(A.16)}\\
(\bar\p\cdot\bar\p)\bqexp(\alpha)&=&-\alpha
Q_N\bqexp(\alpha)+q^{-N}\alpha^2(x\cdot x)\bqexp(\alpha)+GT\ .
\label{(A.17)}
\end{eqnarray}

Thus we have four types of the Laplacians which have the eigenfunction
given by the $q$-exponential functions:
\begin{eqnarray}
\ [-q^{-N}(\p\cdot\p)+\alpha^2(x\cdot x)]\qexp(\alpha)&=&q^{-N}\alpha
Q_N\qexp(\alpha)\ , \label{(A.22)}\\
\ [-q^N(\bar\p\cdot\bar\p)+\alpha^2(x\cdot x)]\bqexp(\alpha)&=&q^N\alpha
Q_N\bqexp(\alpha)\ , \label{(A.23)}\\
\ [-q^{-N}\lambda(\p\cdot\bar\p)+\alpha^2(x\cdot
x)]\qexp(\alpha)&=&q^{-N}\alpha Q_N\qexp(\alpha)\ , \label{(A.24)}\\
\ [-q^N\lambda^{-1}(\bar\p\cdot\p)+\alpha^2(x\cdot
x)]\bqexp(\alpha)&=&q^N\alpha Q_N\bqexp(\alpha)\ . \label{(A.25)}
\end{eqnarray}

\end{document}